\documentclass[a4paper,12pt]{article}
\usepackage{graphics}
\usepackage[utf8]{inputenc}
\usepackage{graphicx}
\usepackage{amsmath}
\usepackage{amsfonts}
\usepackage{amssymb}
\usepackage[english]{babel}
\usepackage{url}
\usepackage[colorlinks=true,urlcolor=blue,citecolor=blue]{hyperref}
\usepackage[numbers,sort&compress]{natbib}
\usepackage{tensor}
\providecommand{\U}[1]{\protect\rule{.1in}{.1in}}
\providecommand{\U}[1]{\protect\rule{.1in}{.1in}}
\begin{document}
	
	\title{\LARGE\bf Influence of a cosmic string on the rate of pairs produced by the Coulomb potential}
	\author{{\small B. Belbaki}\thanks{E-mail: b.belbaki@univ-jijel.dz} {\ \small and} 
		{\small A. Bounames}\thanks{E-mail: bounames@univ-jijel.dz} \\
		{\small Laboratory of Theoretical Physics, Department of Physics}, \\ {\small University of Jijel, Jijel 18000, Algeria.}}
	\date{}
	\maketitle
	
	\begin{abstract}
		We study particle creation phenomenon by the Coulomb potential of an external electric field in the presence of a gravitational field of a static cosmic string. For that, the generalized Klein-Gordon and Dirac equations are solved, and by using the Bogoliubov transformation we calculate the probability and the number density of created particles. It is shown that the presence of the cosmic string enhances the particle production. For the grand unified theory (GUT) cosmic string, the production of spinless particles is possible if the Coulomb potential nucleus charge $Z\geq206$, and for spin-1/2 particles if $Z\geq275$.\\\\
		
	\end{abstract}

\newpage

\section{Introduction}
The creation of particles by an electric field in curved spacetime is an
active research field and continue to attract attentions as in the
cosmological models of an expanding universe \cite{davies,hawking,bernard,lapedes,Buchbinder,mottola,lotze,garriga,Gavrilov,greiner2,mendy,merad,chekireb,chekireb1,haouat,sogut,rajeev,hamil,pimentel}. In particular, the creation of particles by the Coulomb potential of an
external electric field is a topic that has received a special interest
\cite{lapedes,pieper,gershtein,popov1,rafelski1,soffel,baloi} and also in the references cited in the three relevant reviews \cite{popov2,rafelski,voskresensky}. Moreover, particle
production under the effect of topological defects has been also discussed in
the literature, like in the field of a magnetic monopole and domain walls
\cite{pullin,duru}, and in the presence of a cosmic string
\cite{parker3,sahni,mendell,harari,garriga2,jensen,brevik,lorenci1,lorenci2,bezerra1}.

The cosmic strings are hypothetical objects which may have been formed during
the inflationary phase in the primordial universe
\cite{kibble,zeldovich,vilenkin}. They are one-dimensional topological defects
in the spacetime structure and resulting from a symmetry breaking at an energy
close to $10^{16}$ $GeV$ ($10^{-36}$ seconds) \cite{vilenkin}, and with a
maximum mass per unit length $\mu_{\max}=6.73\times10^{27}g.cm^{-1}$
\cite{gott}. They have a nontrivial topology where the spacetime is locally
flat and globally conical with an azimuthal deficit angle \cite{linet1}. A cosmic
string induces a repulsive force on an electric charge at rest \cite{linet2} or on a current \cite{furtado}, 
and an attractive force on a neutral particle \cite{smith}. It has also relevant effects 
like gravitational lensing \cite{vilenkin}, Casimir effect and the gravitational Aharonov-Bohm effect \cite{dowker1,dowker2}. The various observation programs of the anisotropies in cosmic background radiation (CMB) by COBE, WMAP and the Planck satellite have not observed any
cosmic strings effects on the primordial density perturbations \cite{planck}.
Nevertheless, it is possible that they have a role in the production of
gravitational waves \cite{Sakellariadou2,damour2,blasi,boileau} and generation
of high-energy cosmic rays \cite{battacharjee}. More recently, a two-level
static atom coupled to an electromagnetic field, in a cosmic string spacetime,
is suggested as a detector to estimate the deficit angle \cite{yang}.

In order to introduce the deficit angle, let us use the metric of an infinite straight string in cylindrical coordinates $(t,\rho,\phi,z)$ \cite{vilenkin,anderson} 
\begin{equation}
	ds^{2}=-dt^{2}+d\rho^{2}+\alpha^{2}\rho^{2}d\phi^{2}+dz^{2}, \label{111}%
\end{equation}
where the cosmic string parameter $\alpha=1-4\bar{\mu}$ varies in the interval $]0,1]$ with $\bar{\mu}=\frac{G}{c^2}\mu$, $\mu$ the linear mass density of the string, $G$ the gravitational constant, $c$ the speed of light and the variation of the angular variable is $0\leq\phi\leq2\pi$.
The metric (\ref{111}) has been obtained by solving Einstein's equations \cite{vilenkin,anderson}. If we introduce the change of variable $\varphi=(1-4\bar{\mu})\phi$, the metric (\ref{111}) reduces to the Minkowski line element 
\[ ds^{2}=-dt^{2}+d\rho^{2}+\rho^{2}d\varphi^{2}+dz^{2},\]
where the interval of variation of $\varphi$ is
\begin{equation}
	0\leq\varphi\leq2\pi(1-4\bar{\mu}),
\end{equation}
and consequently the cosmic string spacetime is locally flat and globally conical
with a deficit angle $\Delta=8\pi\bar{\mu}=2\pi(1-\alpha)$, i.e., its geometry has a
conical singularity and the curvature tensor is defined using the 2D Dirac
delta function $\delta^{(2)}(\rho)$ \cite{delta}
\begin{equation}
	R\indices{^{12}_{12}}=2\pi\frac{\alpha-1}{\alpha}\delta^{(2)}(\rho),
\end{equation}
which means that the curvature is concentrated on the cosmic string axis and
zero outside. In the absence of the string $\alpha=1$, the curvature tensor vanishes. The case 
$\alpha>1$ corresponds to an anti-conical spacetime with negative curvature \cite{katanaev,furtado1}.

Thereby, it is physically meaningful to analyze the role of cosmic string
deficit angle on the rate of pairs produced by an external electric field. For
this purpose, in this paper we study particle creation by a vector Coulomb
potential in the presence of a static cosmic string. Once the solutions of the
Klein-Gordon and Dirac equations obtained, the expressions of the probability
and the number density of the created particles will be computed using the
Bogoliubov transformation. Thereafter, we discuss the influence of the cosmic
string on the rate of created particles induced by the Coulomb potential for
both spin-$0$ boson and spin$-1/2$ fermion particles. On the other hand, it is
worth mentioning that the self-adjoint extensions method has recently been
used to construct rigorous mathematical formulation of wave equation solutions
for the Coulomb potential and similar singular potentials \cite{gitman1,wiese}.

In this work, we will use spherical coordinates since the problem has
spherical symmetry, for that we consider the coordinate transformation
$\rho=r\sin\theta$ and $z=r\cos\theta$, and the line element of the cosmic
string becomes
\begin{equation}
	ds^{2}=-dt^{2}+dr^{2}+r^{2}d\theta^{2}+\alpha^{2}r^{2}\sin^{2}\theta d\phi^{2}, \label{1}%
\end{equation}
where $(t,r,\theta,\phi)$: $-\infty<t<+\infty$, $r\geq0$, $0\leq\theta
\leq\pi/2$ \ and $0\leq\phi\leq2\pi$. In the absence of the string
$\alpha=1$, the metric (\ref{1}) reduces to the Minkowski one. \newline
The present paper is organized as follows, in sections 2 and 3 we present
respectively the solutions of the Klein-Gordon and Dirac equations in the
presence of a Coulomb potentiel in a cosmic string spacetime, for both cases
we calculate the probability and the number density of created particles. The 
last section is devoted to the conclusion.

\section{Klein-Gordon equation in cosmic string space-time}
In curved spacetime, the generalized Klein-Gordon equation of spin$-0$
particle of mass $M$ and charge $q$ minimally coupled to an external
electromagnetic field $A_{\mu}$ is
\begin{equation}
	\left[  {-\frac{1}{\sqrt{-g}}D_{\mu}{g^{\mu\nu}}\sqrt{-g}D_{\nu}+{M^{2}}}
	\right]  \Psi=0, \label{kg}%
\end{equation}
where $D_{\mu}={\partial_{\mu}}+iq{A}_{\mu}$ is the covariant derivative,
${{g^{\mu\nu}}}$ the metric tensor of the curved spacetime and $g=\det({{
		g^{\mu\nu}}})$. \newline In this work, we consider the case where the spacial
components of the external field have a null value $({A}_{i}=0;i=1,2,3)$ and
its time component $A_{0}$ is the Coulomb potential generated by a point-like
source charge $Q_{s}$ where its expression is
\begin{equation}
	{A_{0}}\equiv{V(r)}=\frac{Q_{s}}{r}, \label{6}%
\end{equation}
and for an electron in a hydrogen-like atom $Q_{s}=Ze$ and $q=-e$.\newline
Thus, the Klein-Gordon equation (\ref{kg}) can be written explicitely in the
cosmic string metric like 
{\footnotesize 
\begin{equation}
	\left[  {-{{\left(  {\frac{\partial}{{\partial t}}+iq{A_{0}}}\right)  }^{2}}+
		\frac{1}{{{r^{2}}}}\frac{\partial}{{\partial r}}\left(  {{r^{2}}\frac{
				\partial}{{\partial r}}}\right)  +\frac{1}{{{r^{2}}\sin\theta}}\left(  {
			\cos\theta\frac{\partial}{{\partial\theta}}+\sin\theta\frac{{{ \partial^{2}}}%
			}{{\partial{\theta^{2}}}}}\right)  +\frac{1}{{{\alpha^{2}}{r^{2}}{{\sin}^{2}%
				}\theta}}\frac{{{\partial^{2}}}}{{\partial{\phi^{2}}} }-{M^{2}}}\right]
	\Psi=0. \label{2}%
\end{equation}
}
Since the potential (\ref{6}) is time-independent and the problem has a
spherical symmetry, the solution can be taken in the form
\begin{equation}
	\Psi(t,r,\theta,\phi)=\frac{{u(r)}}{r}f(\theta){e^{i(-Et+m\phi)}},
	\label{3}%
\end{equation}
where $m=0,\pm1,\pm2,...$ and $E$ is the energy of the spin-$0$ particle.
\newline By Substituting Eq. (\ref{3}) in Eq. (\ref{2}), the angular function
$f(\theta)$ satisfies the following differential equation
\begin{equation}
	\left[  {\frac{1}{{\sin\theta}}\frac{d}{{d\theta}}\left(  {\sin\theta\frac
			{d}{{d\theta}}}\right)  -\frac{{{m^{2}}}}{{{\alpha^{2}}{{\sin}^{2}} \theta}}%
	}\right]  f(\theta)=-{\lambda_{\alpha}}f(\theta), \label{4}%
\end{equation}
where its solution is given in terms of the generalized Legendre functions $P_{\nu}^{\mu}(x)$, with
\begin{equation}
	{\lambda_{\alpha}}={l_{\alpha}}({l_{\alpha}}+1),
\end{equation}
where ${l_{\alpha}}=n+\left\vert {{ m_{\alpha}}}\right\vert =l+\left\vert m\right\vert \left(\frac{1}{\alpha}-1\right)$ with $n$ a non-negative integer, $l=n+ \left\vert m\right\vert$ and $m_{\alpha}={\frac{m}{\alpha}}$. In addition $m_{\alpha}$ and $l_{\alpha}$ are not necessarily integers, and $m_{\alpha}$ varies in the range $-{l_{\alpha}}\leq{m_{\alpha}}\leq{l_{\alpha}}$. Finally, $l$ and $m$ are the
orbital angular momentum and the magnetic quantum numbers in the flat space, respectively \cite{santos}.

On the other hand, the radial function $u(r)$ satisfies the following
differential equation \cite{santos}
\begin{equation}
	\frac{{{d^{2}}u}}{{d{r^{2}}}}+\left[  {{K^{2}}-{V_{eff}}-\frac{{{l_{\alpha} }(
				{l_{\alpha}}+1)}}{{{r^{2}}}}}\right]  u=0, \label{5}%
\end{equation}
where $K^{2}$ and $V_{eff}$ are defined as
\begin{equation}
	{K^{2}}={E^{2}}-{M^{2}}, \qquad  {V_{eff}}=2qE{V(r)}-q^{2}{V}^{2}(r).
\end{equation}
By substitution of the expression (\ref{6}) of the potential, Eq. (\ref{5})
can be written in the form
\begin{equation}
	\frac{{{d^{2}}u}}{{d{r^{2}}}}+\left[  {{K^{2}}-\frac{2EV_{0}}{r}-\frac{{{
					l_{\alpha}}({l_{\alpha}}+1)}-V_{0}^{2}}{{{r^{2}}}}}\right]  u=0, \label{7}%
\end{equation}
with ${V_{0}}\equiv{qQ_{s}}$. If we study only the case $V_{0}>0$, the results
for the case of $V_{0}<0$ can be obtained by the charge conjugation
transformation \cite{gitman1}.\newline In order to search the solutions
corresponding to the condition $\left\vert E\right\vert >M$, we use the
following notations
\begin{equation}
	{\gamma_{l}}=\pm\sqrt{{{\left(  {{l_{\alpha}}+\frac{1}{2}}\right)  }^{2}}
		-V_{0}^{2}}, \qquad  \eta=\frac{{E}V_{0}}{K}, \label{41}%
\end{equation}
and the change $z=-2iKr$ transforms Eq. (\ref{7}) into the following form
\begin{equation}
	\frac{{{d^{2}}u}}{{d{z^{2}}}}+\left[  {-\frac{1}{4}-\frac{{i\eta}}{z}+\frac{
			\frac{1}{4}{-\gamma_{l}^{2}}}{z{{^{2}}}}}\right]  u=0, \label{8}%
\end{equation}
which admits two regular linearly independent solutions in terms of the
Whittaker functions ${M_{-i\eta,{\gamma_{l}}}}(z)$ and ${W_{-i\eta,{\gamma
			_{l}}}}(z)$ \cite{22}%
\begin{equation}
	u_{1}(z)=C_{1}{M_{-i\eta,{\gamma_{l}}}}(z)=C_{1}{z^{({\gamma_{l}+}\frac{1}{2}
			)}\exp(-z/2)}M({\gamma_{l}}+i\eta+\frac{1}{2},1+2{\gamma_{l}};z), \label{nv1}%
\end{equation}
\begin{equation}
	u_{2}(z)=C_{2}{W_{-i\eta,{\gamma_{l}}}}(z)=C_{2}{z^{({\gamma_{l}+}\frac{1}{2}
			)}\exp(-z/2)}U({\gamma_{l}}+i\eta+\frac{1}{2},1+2{\gamma_{l}};z), \label{nv2}%
\end{equation}
$C_{1}$ and $C_{2}$ are the normalization constant, $M(a,b,z)$ and $U(a,b,z)$
are the Kummer's confluent hypergeometric functions. The first solution
(\ref{nv1}) is bounded at $z=0$ while the second (\ref{nv2}) is bounded at $\vert z \vert \rightarrow\infty$.

\subsection{Creation of scalar particles}

In order to obtain the expressions of the probability and the number density of created particles, we use the Bogoliubov transformation which links the asymptotic behavior of the obtained solutions. 

Indeed, the asymptotic behavior of ${W_{-k,{\mu}}}(z)$ for $\vert z \vert \rightarrow\infty$ is
\begin{equation}
	W_{-k,{\mu}}(z) \rightarrow {z^{-k}\exp(-z/2)}, \label{f18} 
\end{equation}
and the states $\phi^{+}_{out}$ and $\phi^{-}_{out}$ can be defined as
\begin{equation}
\phi^{+}_{out}=C^{+}_{2 out} W_{-k,{\mu}}(z), \quad
\phi^{-}_{out}=[C^{+}_{2 out} W_{-k,{\mu}}(z)]^{*}=[C^{+}_{2 out}]^{*}  W_{k,{\mu}}(-z). \label{f19} 	
\end{equation}	
Similarly, the asymptotic behavior of ${M_{-k,{\mu}}}(z)$ for $z \rightarrow 0$ is
\begin{equation}
	M_{-k,{\mu}}(z) \rightarrow {z^{({\mu+}\frac{1}{2})}}, \label{f20} 	
\end{equation}
and the states $\phi^{+}_{in}$ and $\phi^{-}_{in}$ are 
\begin{equation}
	\phi^{+}_{in}=C^{+}_{1 in} M_{-k,{\mu}}(z), \quad
	\phi^{-}_{in}=[C^{+}_{1 in} M_{-k,{\mu}}(z)]^{*}=[C^{+}_{1 in}]^{*} (-1)^ {(-\mu+1/2)}M_{-k,{-\mu}}(z). \label{f21} 	
\end{equation}
Therefore, the Bogoliubov transformation can be written as
\begin{equation}
	\phi^{+}_{out}(z)={A}\phi^{+}_{in}(z)+B\phi^{-}_{in}(z), \label{nv3}%
\end{equation}
and with the help of the two following functional relations \cite{22}
\begin{equation}
	{W_{-k,\mu}}(z)=\frac{{\Gamma(-2\mu)}}{{\Gamma(-\mu+k+\frac{1}{2})}}{M_{-k,\mu}}(z)+\frac{{\Gamma(2\mu)}}{{\Gamma(\mu+k+\frac{1}{2})}}M_{-k,-\mu
	}(z),\label{nv4}%
\end{equation}
\begin{equation}
	M_{-k,-\mu}(z)={e^{i\pi(\mu-1/2)}} [{{M_{k,\mu}}(z)}]^{*}, \label{224}%
\end{equation}
the Bogoliubov coefficients $A$ and $B$ satisfy
\begin{equation}
	\left\vert {\frac{B}{A}}\right\vert =\left\vert {\frac{{\Gamma(-{{{\gamma}%
						_{l}+i}\eta+{\frac{1}{2}})}}}{{\Gamma({{{\gamma}_{l}+i}\eta+{\frac{1}{2}})}}%
		}{e^{i\pi({\gamma_{l}}-1/2)}}}\right\vert ,
\end{equation}
where ${k={{i}\eta}}$ and ${\mu={{{\gamma}_{l}}}}$.

Using the bosonic condition $\left\vert A\right\vert ^{2}-\left\vert
B\right\vert ^{2}=1$ and $\gamma_{l}=i{\tilde{\gamma}_{l},}$ the probability
for one pair production is
\begin{equation}
	P={\left\vert {\frac{B}{A}}\right\vert ^{2}}={\left\vert \frac{{\Gamma
				(-i\tilde{\gamma}_{l}+{{i}\eta+{\frac{1}{2}})}}}{{\Gamma(i\tilde{\gamma}%
				_{l}{{+i}\eta+{\frac{1}{2}})}}}{{e^{i\pi({i\tilde{\gamma}_{l}}-1/2)}}%
		}\right\vert ^{2}.} \label{223}%
\end{equation}
Then, using the property of gamma function
\begin{equation}
	{\left\vert {\Gamma\left(  {\frac{1}{2}+ix}\right)  }\right\vert ^{2}}%
	=\frac{\pi}{{\cosh(\pi x)}},
\end{equation}
the expression of the probability is reduced to
\begin{equation}
	P=\frac{\cosh\left[  \pi(\tilde{\gamma}_{l}+\eta)\right]  }{\cosh\left[
		\pi(\tilde{\gamma}_{l}-\eta)\right]  }e^{-2\pi\tilde{\gamma}_{l}}, \label{222}%
\end{equation}
where
\begin{equation}
	{\tilde{\gamma}_{l}}=\sqrt{V_{0}^{2}-\left[  l+\left\vert m\right\vert \left(
		\frac{1}{\alpha}-1\right)  +\frac{1}{2}\right]  ^{2}}>0, \label{cond1}%
\end{equation}
is the condition for pair production of spinless particle which depends on
${V_{0}}={qQ_{s}}$, the cosmic string paramter $\alpha$ and the two quantum
numbers $l$ and $m$.

The number density of the scalar particles created is given by
\begin{equation}
	\bar{n}=\left\vert B\right\vert ^{2}=\frac{1}{P^{-1}-1}=\frac{\cosh\left[
		\pi(\tilde{\gamma}_{l}+\eta)\right]  }{\cosh\left[  \pi(\tilde{\gamma}%
		_{l}-\eta)\right]  e^{2\pi\tilde{\gamma}_{l}}-\cosh\left[  \pi(\tilde{\gamma
		}_{l}+\eta)\right]  }, \label{466}%
\end{equation}
which is positive since ${{\tilde{\gamma}}_{l}}>0$. 

At large frequencies $(E\gg M)$, the probability (\ref{222}) reduces to
\begin{equation}
	P=\frac{\cosh\left[  \pi(\tilde{\gamma}_{l}+V_{0})\right]  }{\cosh\left[
		\pi(\tilde{\gamma}_{l}-V_{0})\right]  }e^{-2\pi\tilde{\gamma}_{l}},
	\label{220}%
\end{equation}
and the number density takes the form
\begin{equation}
	\bar{n}=\frac{\cosh\left[  \pi(\tilde{\gamma}_{l}+V_{0})\right]  }%
	{\cosh\left[  \pi(\tilde{\gamma}_{l}-V_{0})\right]  e^{2\pi\tilde{\gamma}_{l}%
		}-\cosh\left[  \pi(\tilde{\gamma}_{l}+V_{0})\right]  }, \label{47}%
\end{equation}
which is not a thermal distribution.

For the case when  $\left\vert\tilde{\gamma_{l}}\pm V_{0}\right\vert\gg1$, the number
density (\ref{47}) takes the form of Bose-Einstein distribution
\begin{equation}
	\bar{n}=\frac{1}{e^{2\pi(\tilde{\gamma}_{l}-V_{0})}-1}. \label{48}%
\end{equation}
In the absence of the cosmic string ($\alpha=1)$, or for $m=0$, the
probability (\ref{220}) and the number density (\ref{47}) reduce to
\begin{equation}
	P(\alpha=1)=\frac{\cosh\left[  \pi(\tilde{\gamma}_{1}+V_{0})\right]  }%
	{\cosh\left[  \pi(\tilde{\gamma}_{1}-V_{0})\right]  }\exp\left[  -2\pi
	\tilde{\gamma}_{1}\right],
\end{equation}
\begin{equation}
	\bar{n}(\alpha=1)=\frac{\cosh\left[  \pi(\tilde{\gamma}_{1}+V_{0})\right]
	}{\cosh\left[  \pi(\tilde{\gamma}_{1}-V_{0})\right]  e^{2\pi\tilde{\gamma}
			_{1}}-\cosh\left[  \pi(\tilde{\gamma}_{1}+V_{0})\right]  }, \label{4661}%
\end{equation}
where $\tilde{\gamma}_{1}\equiv\tilde{\gamma}_{l}(\alpha=1)= \sqrt{V_{0}^{2}-{{\left({l+\frac{1}{2}}\right)}^{2}}}$.

\subsection{Discussion of results}
$\bullet$ First, let us discuss the consequences of the condition
$(\ref{cond1})$, for particle production of spin-0 boson, which can be written
as
\begin{equation}
	\left\vert qQ_{s}\right\vert > \left[l+\left\vert m\right\vert \left(  \frac{1}%
	{\alpha}-1\right)  +\frac{1}{2}\right]  . \label{cond11}%
\end{equation}
In the absence of the cosmic string ($\alpha=1)$, or for $m=0$, the above
condition is reduced to
\begin{equation}
	\left\vert qQ_{s}\right\vert > \left(l+\frac{1}{2}\right),
\end{equation}
which is similar to that obtained in Eq. (37) for pair creation in a magnetic
monopole field \cite{duru}, and also similar to the results discussed after
Eq. (84) for pair creation by charged black holes \cite{kim}.

For an electron in a hydrogen-like atom, $Q_{s}=Ze$ and $q=-e$, the condition (\ref{cond11}) can be expressed for the nucleus charge $Z$ as
\begin{equation}
	Z>Z_{cr}\equiv\frac{1}{e^{2}}\left[  l+\left\vert m\right\vert (\frac{1}{\alpha
	}-1)+\frac{1}{2}\right]  .
\end{equation}
For the particular values of the quantum numbers $(l=0,m=0)$, the pair
production of scalar particles is possible for : $Z>Z_{cr}=\frac{1}{2e^{2}%
}=\frac{137}{2}$ (in units $\hbar=c=1$, $e^{2}\simeq1/137$), which is a well-known result in the absence of the cosmic string \cite{popov1}.

In table \ref{table1}, for the first values of the quantum numbers $(l,m),$
we give the critical values $Z_{cr}$ as function of the cosmic string
parameter $\alpha$ and an estimation of the effect of the GUT cosmic string (for which \cite{marques}: $\alpha=1-10^{-6}=0.999999$) on the values and the rate of $Z_{cr}^{GUT}$.

\begin{table}[h]
\begin{center}
	\tabcolsep=21pt
	\renewcommand\arraystretch{1.2}
	\begin{minipage}{\textwidth}
	\caption{$Z_{cr}$ in terms of $\alpha$, values and the rate of $Z_{cr}^{GUT}$ for boson particle} \label{table1}
	\begin{tabular}
		[c]{|c|c|c|c|c|}\hline
		$l$ & $m$ & $Z_{cr}$ & $Z_{cr}^{GUT}$ & $\frac{Z_{cr}%
			^{GUT}-Z_{cr}(\alpha=1)}{Z_{cr}(\alpha=1)}$\\\hline
		$0$ & $0$ & $\frac{137}{2}=68.5$ &  & \\\hline
		$1$ & $0$ & $137\times\frac{3}{2}=205.5$ &  & \\\cline{2-5}
		& $\pm1$ & $137\times\left(  \frac{1}{\alpha}+\frac{1}{2}\right)  $ &
		$205.500137$ & $6.666666\times10^{-5}\%$\\\hline
		& $0$ & $137\times\frac{5}{2}=342.5$ &  & \\\cline{2-5}%
		$2$ & $\pm1$ & $137\times\left(  \frac{1}{\alpha}+\frac{3}{2}\right)  $ &
		$342.500137$ & $4.000004\times10^{-5}\%$\\\cline{2-5}
		& $\pm2$ & $137\times\left(  \frac{2}{\alpha}+\frac{1}{2}\right)  $ &
		$342.500274$ & $8\times10^{-5}\%$\\\hline
	\end{tabular}
\end{minipage}
\end{center}
\end{table}
\vspace*{1mm}

We note that for the sub-states with $m=0$, the critical value $Z_{cr}$
increases with $l$ and is independent of the cosmic string parameter $\alpha$
(Note that these sub-states are equivalent to the case $\alpha=1$). For the
sub-states for which $m\neq0$, $Z_{cr}$ increases with $l$ and depends
inversely on the cosmic string parameter $\alpha$. For the GUT
cosmic string, the obtained critical values $Z_{cr}^{GUT}$ do not exist for the moment (at present, the maximum is $Z=118$ in Mendeleev table), the rate of $Z_{cr}^{GUT}$ is about $10^{-5}\%$ as compared with the case $\alpha=1$
(without cosmic string) and the production of scalar particles is possible if the Coulomb potential nucleus charge (or atomic number) $Z\geq206$.\\

\begin{figure}[h]
	\begin{center}%
		\begin{tabular}
			[c]{cc}%
			\includegraphics[width=0.55\textwidth]{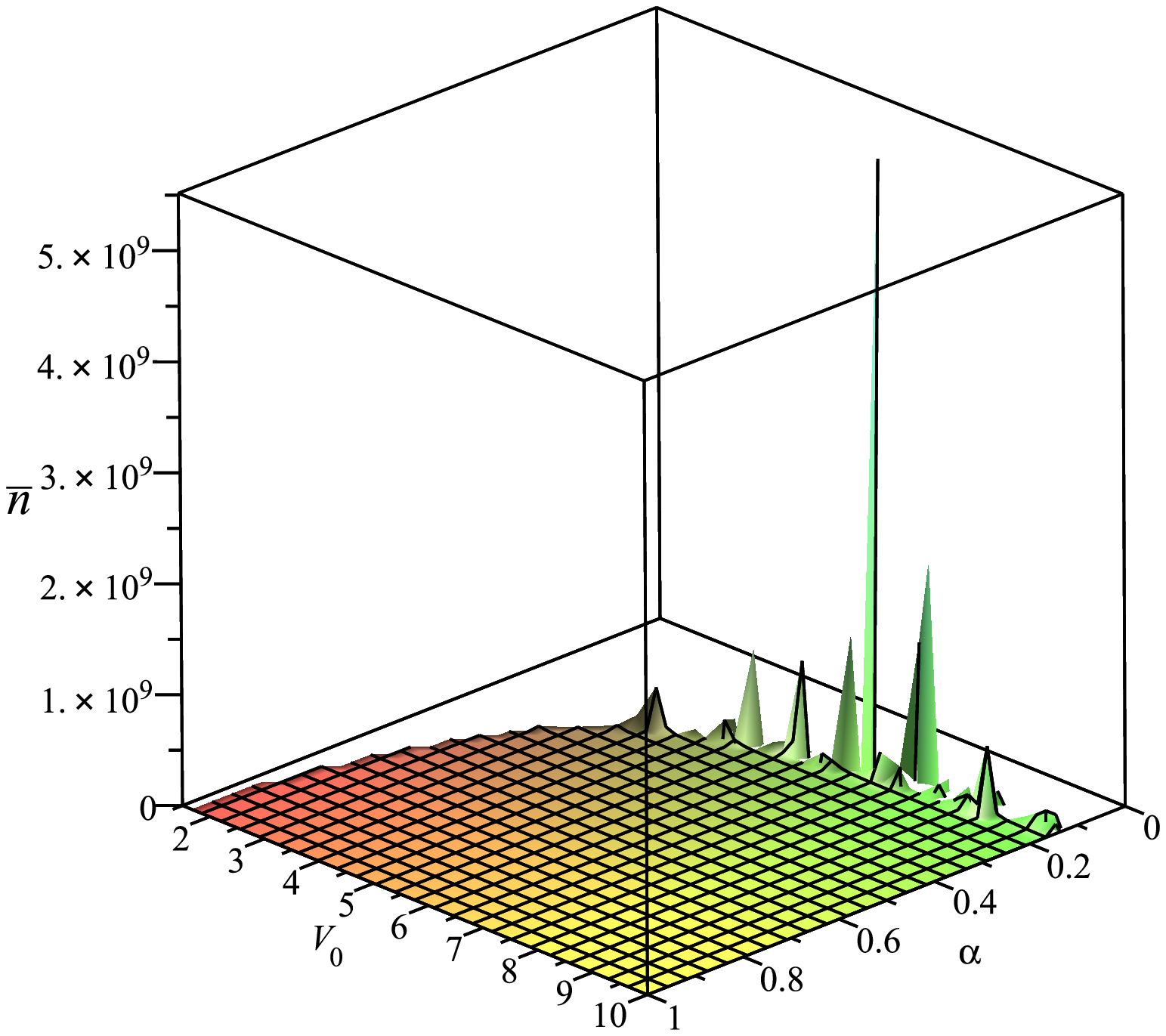} &
			\includegraphics[width=0.4\textwidth]{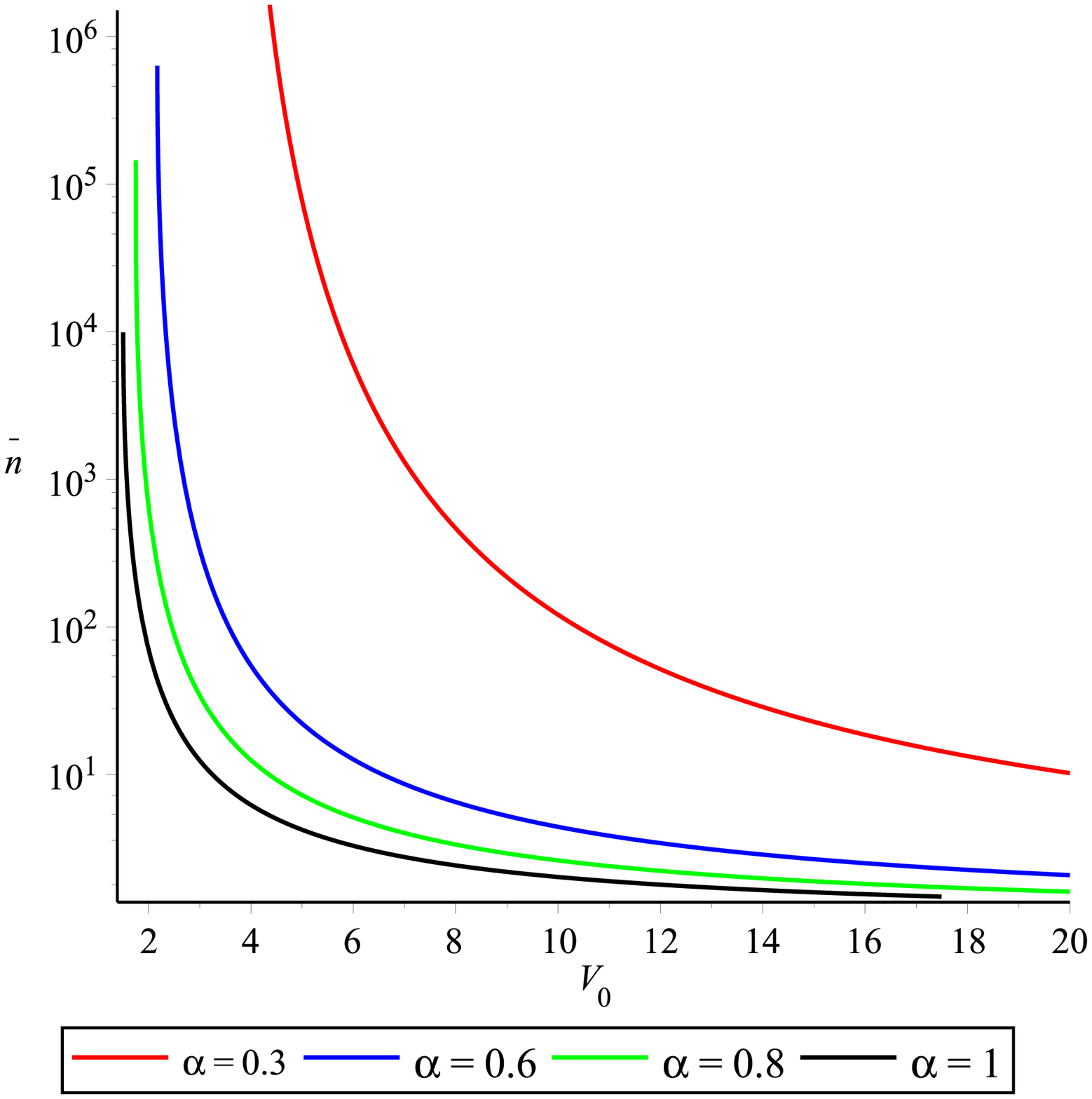}\\
			(a) & (b)
		\end{tabular}
	\end{center}
	\begin{center}
		\caption{Plot (a) is the 3D number density (\ref{47}) in terms of $V_{0}$ and $\alpha$. Plot (b) is the 2D number density (\ref{47}) in terms of $V_{0}$ for different values of $\alpha=0.3, 0.6, 0.8$ and $1$ (without cosmic string) \label{fig1}.}	
	\end{center}
\end{figure}

$\bullet$ Secondly, in Figure $\ref{fig1}$ the number density curves have been plotted
for fixed values of the quantum numbers $(l=1,m=1)$ and where the condition $(\ref{cond1})$ for pair production of 
spinless particles is reduced to $ V_{0} > (\frac{1}{2}+\frac{1}{\alpha})$ which should be always fullfilled. Plot (a) display, in three (3D) dimensions, 
the number density $(\ref{47})$ in terms of $V_{0}=qQ_{s}$ and the cosmic string parameter $\alpha$. Plot (b) display, in two (2D)
dimensions, the number density $(\ref{47})$ in terms of $V_{0}=qQ_{s}$ for different values the cosmic string parameter $\alpha$.\\
 
From plot (b) we note that the number density $\bar{n}$ decreases when the cosmic string parameter $\alpha$ increases (i.e. the linear mass density $\mu$ of the string decreases) where its smallest values correspond to the case $\alpha = 1$ (in the absence of the cosmic string). Therefore, we deduce that the presence of the cosmic string improves the number density of scalar particles compared to the case without the cosmic string ($\alpha=1$).


\section{Dirac equation in cosmic string space-time}

In curved spacetime, the generalized Dirac equation of spin$-1/2$ particle of
mass $M$ and charge $q$ minimally coupled to an external electromagnetic field
$A_{\mu}$ is%
\begin{equation}
	\left[  {i{\gamma^{\mu}}(x)\left(  {{\nabla_{\mu}}+i{qA_{\mu}}}\right)
		-M}\right]  \Psi(x)=0, \label{15}%
\end{equation}
where ${\nabla_{\mu}}={\partial_{\mu}}+{\Gamma_{\mu}}(x)$ is the covariant
derivative, $\Gamma_{\mu}$ is the spinor affine connection
\begin{equation}
	{\Gamma_{\mu}}=\frac{1}{4}{\gamma^{(a)}}{\gamma^{(b)}}{e^{\nu}}_{(a)}\left[
	{{\partial_{\mu}}{e_{(b)\nu}}-\Gamma_{\mu\nu}^{a}{e_{(b)\sigma}}}\right]  ,
\end{equation}
$\Gamma_{\mu\nu}^{a}$ is the Christoffel symbol of the second kind,
$\{{e^{\mu}}_{(a)}(x)\}$ the tetrad basis and ${{\gamma^{\mu}}(x)}$ the
generalized Dirac matrix satisfying the relation
\begin{equation}
	\left\{  {{\gamma^{\mu}}(x),{\gamma^{\nu}}(x)}\right\}  =2{g^{\mu\nu}}(x),
\end{equation}
which can be written as a function of the tetrad basis $\{{e^{\mu}}%
_{(a)}(x)\}$ and the standard flat spacetime Dirac matrices ${\gamma_{(a)}}$
as
\begin{equation}
	{\gamma^{\mu}}(x)={e^{\mu}}_{(a)}(x){\gamma_{(a)}}. \label{12}%
\end{equation}
The tetrad basis $\{{e^{\mu}}_{(a)}(x)\}$ satisfies the relations
\begin{equation}
	{\eta^{(a)(b)}}{e^{\mu}}_{(a)}(x){e^{\nu}}_{(b)}(x)={g^{\mu\nu}}(x),
\end{equation}
where the Greek letters are for tensor indices and Latin letters for tetrad indices.

We consider the case where the spacial components of the external field are
nuls $({A}_{i}=0,i=1,2,3)$ and its time component ${A_{0}}(r) \equiv{V}(r)$ is
the Coulomb potential defined in Eq. (\ref{6}).

In order to write the Dirac equation in the cosmic string metric (\ref{1}), we
use the associated tetrad $e_{(a)}^{\mu}(x)$ defined by
\cite{marques,hassanabadi}%

\begin{equation}
	{e^{\mu}}_{(a)}=\left(  {
		\begin{array}
			[c]{cccc}%
			1 & 0 & 0 & 0\\
			0 & {\sin\theta\cos\phi} & {\sin\theta\sin\phi} & {\cos\theta}\\
			0 & {\frac{{\cos\theta\cos\phi}}{r}} & {\frac{{\cos\theta\sin\phi}}{r}} & {-
				\frac{{\sin\theta}}{r}}\\
			0 & {-\frac{{\sin\phi}}{{\alpha r\sin\theta}}} & {\frac{{\cos\phi}}{{\alpha
						r\sin\theta}}} & 0
		\end{array}
	}\right)  , \label{13}%
\end{equation}
and the expressions of the generalized Dirac matrices $\gamma^{\mu}(x)$ are
\begin{equation}
	{\gamma^{0}}(x)={\gamma^{(0)}},
\end{equation}%
\begin{equation}
	{\gamma^{1}}(x)=\sin\theta\cos\phi{\gamma^{(1)}}+\sin\theta\sin\phi{
		\gamma^{(2)}}+\cos\theta{\gamma^{(3)}},
\end{equation}%
\begin{equation}
	{\gamma^{2}}(x)=\frac{{\cos\theta\cos\phi}}{r}{\gamma^{(1)}}+\frac{{\cos
			\theta\sin\phi}}{r}{\gamma^{(2)}}-\frac{{\sin\theta}}{r}{\gamma^{(3)}},
\end{equation}%
\begin{equation}
	{\gamma^{2}}(x)=-\frac{{\sin\phi}}{{\alpha r\sin\theta}}{\gamma^{(1)}}+\frac{
		{\cos\phi}}{{\alpha r\sin\theta}}{\gamma^{(2)}}.
\end{equation}
Since the problem is time-independent, the wavefunction can be written as
\begin{equation}
	\Psi(x)={e^{-iEt}}\chi(\vec{r}), \label{14}%
\end{equation}
and by substitution of Eq. (\ref{14}) in Eq. (\ref{15}), we obtain \cite{marques}
{\footnotesize 
\begin{equation}
	\left[  {i{\gamma^{(r)}}{\partial_{r}}+i\frac{{{\gamma^{(\theta)}}}}{r}{
			\partial_{\theta}}+i\frac{{{\gamma^{(\phi)}}}}{{\alpha r\sin\theta}}{
			\partial_{\phi}}+{\gamma^{(0)}}E+i\frac{1}{{2r}}\left(  {1-\frac{1}{\alpha}}
		\right)  \left(  {{\gamma^{(r)}}+\cot\theta{\gamma^{(\theta)}}}\right)  -{
			\gamma^{(0)}}{qA_{0}}-M}\right]  \chi(\vec{r})=0, \label{16}%
\end{equation}
}
where
\begin{equation}
	\left(  {
		\begin{array}
			[c]{c}%
			{{\gamma^{(r)}}}\\
			{{\gamma^{(\theta)}}}\\
			{{\gamma^{(\phi)}}}%
		\end{array}
	}\right)  =\left(  {
		\begin{array}
			[c]{ccc}%
			{\sin\theta\cos\phi} & {\sin\theta\sin\phi} & {\cos\phi}\\
			{\cos\theta\cos\phi} & {\cos\theta\sin\phi} & {-\sin\phi}\\
			{-\sin\phi} & {\cos\phi} & 0
		\end{array}
	}\right)  \left(  {
		\begin{array}
			[c]{c}%
			{{\gamma^{(1)}}}\\
			{{\gamma^{(2)}}}\\
			{{\gamma^{(3)}}}%
		\end{array}
	}\right)  .
\end{equation}
Then, since the problem has a spherical symmetry, the solution of Eq.
(\ref{16}) can be taken as
\begin{equation}
	\chi(\vec{r})={r^{-\frac{1}{2}\left(  {1-\frac{1}{\alpha}}\right)  }}{
		(\sin\theta)^{-\frac{1}{2}\left(  {1-\frac{1}{\alpha}}\right)  }}R(r)Y_{{
			l_{(\alpha)}}}^{{m_{(\alpha)}}}(\theta,\phi), \label{17}%
\end{equation}
where $R(r)$ is the radial part and $Y_{{l_{(\alpha)}}}^{{m_{(\alpha)}}%
}(\theta,\phi)$ the spherical harmonics. The parameters  $m_{\alpha}={\frac{m}{\alpha}}$ and
${l_{\alpha}}=n+\left\vert {{m_{\alpha}}}\right\vert =l+\left\vert m\right\vert \left(
{\frac{1}{\alpha}-1}\right)$ with $l=n+\left\vert m\right\vert$ and $n$ a non-negative integer \cite{marques1}.

By substitution of Eq. (\ref{17}) into Eq. (\ref{16}), we obtain \cite{marques}
\begin{equation}
	\left(  {{\sigma_{y}}{p_{r}}+i\frac{{{\sigma_{y}}}}{r}{\gamma^{(0)} }{
			k_{(\alpha)}}+{qA_{0}}+M{\gamma^{(0)}}}\right)  R(r)=ER(r), \label{18}%
\end{equation}
where $p_{r}$ is the radial momentum operator, ${k_{(\alpha)}}=\pm\left[  {
	j+\left\vert m\right\vert \left(  {\frac{1}{\alpha}-1}\right)  +\frac{1}{2}
}\right]$ with $j$ is the eigenvalue of the total angular momentum operator
in flat Minkowski spacetime \cite{marques1}, and $\sigma_{y}$ is Pauli's matrix
\begin{equation}
	{\sigma_{y}}=\left(  {\
		\begin{array}
			[c]{cc}%
			0 & {-i}\\
			i & 0
		\end{array}
	}\right)  . \label{19}%
\end{equation}
For the radial solution we use the following ansatz \cite{greiner}
\begin{equation}
	R(r)=\frac{1}{r}\left(  {\
		\begin{array}
			[c]{c}%
			{{u_{1}}(r)}\\
			iu{{_{2}}(r)}%
		\end{array}
	}\right)  , \label{20}%
\end{equation}
and by substitution of Eq.(\ref{20}) in Eq. (\ref{18}) and using Eq.
(\ref{19}), we have
\begin{equation}
	\frac{{d{u_{1}}(r)}}{{dr}}=-\frac{{{k_{(\alpha)}}}}{r}{u_{1}}(r)+\left[  {E+M-
		{qA_{0}}}\right]  u{_{2}}(r), \label{22}%
\end{equation}
\begin{equation}
	\frac{{d}u{{_{2}}(r)}}{{dr}}=\frac{{{k_{(\alpha)}}}}{r}u{_{2}}(r)-\left[  {
		E-M-{qA_{0}}}\right]  {u_{1}}(r). \label{21}%
\end{equation}
By taking ${V_{0}}\equiv{qQ_{s}}$ and $A_{0}$ as the Coulomb potential defined
in Eq. (\ref{6}), the above equations become
\begin{equation}
	\frac{{d{u_{1}}(r)}}{{dr}}=-\frac{{{k_{(\alpha)}}}}{r}{u_{1}}(r)+\left[  {E+M-
	}\frac{{V{_{0}}}}{r}\right]  u{_{2}}(r), \label{23}%
\end{equation}
\begin{equation}
	\frac{{d}u{{_{2}}(r)}}{{dr}}=\frac{{{k_{(\alpha)}}}}{r}u{_{2}}(r)-\left[  {
		E-M-}\frac{{V{_{0}}}}{r}\right]  {u_{1}}(r). \label{24}%
\end{equation}
If we study only the case $V_{0}>0$, the results for the case of $V_{0}<0$ can
be obtained by the charge conjugation transformation \cite{gitman1}. Its
convenient to introduce the new coordinate
\begin{equation}
	z=2iKr, \qquad K=\sqrt{E^{2}-M^{2}}, \label{25}%
\end{equation}
then Eqs. (\ref{23})-(\ref{24}) transform into
\begin{equation}
	\frac{{d{u_{1}}(z)}}{{dz}}=-\frac{{{k_{(\alpha)}}}}{z}{u_{1}}(z)+\left[
	\frac{E+M}{2iK}-\frac{V_{0}}{z}\right]  {u_{2}}(z), \label{27}%
\end{equation}
\begin{equation}
	\frac{{d}u{{_{2}}(z)}}{{dz}}=\frac{{{k_{(\alpha)}}}}{z}u{_{2}}(z)-\left[
	\frac{E-M}{2iK}-\frac{V_{0}}{z}\right]  {u_{1}}(z), \label{28}%
\end{equation}
and using the ansatz
\begin{equation}
	u_{1}(z)=\sqrt{E+M}\left[  F(z)+G(z)\right],\\ \quad
	u{_{2}}(z)=i\sqrt{E-M}\left[F(z)-G(z)\right], \label{45}%
\end{equation}
in Eqs. (\ref{27})-(\ref{28}), we have
\begin{equation}
	\frac{{d{F}(z)}}{{dz}}=\left(  \frac{1}{2}-i\frac{V_{0}E}{Kz}\right)  {F}
	(z)+\left(  -\frac{{{k_{(\alpha)}}}}{z}-i\frac{V_{0}M}{Kz}\right)  {G}(z),
	\label{29}%
\end{equation}
\begin{equation}
	\frac{{d{G}(z)}}{{dz}}=\left(  -\frac{1}{2}+i\frac{V_{0}E}{Kz}\right)  {G}
	(z)+\left(  -\frac{{{k_{(\alpha)}}}}{z}+i\frac{V_{0}M}{Kz}\right)  {F}(z)
	\label{30}%
\end{equation}
and from which one easily obtains the following second order differential equation
\begin{equation}
	\frac{d^{2}{F}(z)}{{dz^{2}}}+\frac{1}{z}\frac{{d{F}(z)}}{dz}-\left[  \frac{1
	}{4}+\left(  \frac{1}{2}-i\frac{EV_{0}}{K}\right)  \frac{1}{z}+\frac{
		\gamma_{l}^{2}}{z^{2}}\right]  F(z)=0, \label{31}%
\end{equation}
with $\gamma_{l}^{2}=k_{(\alpha)}^{2}-V_{0}^{2}$.

The following ansatz
\begin{equation}
	M(z)=\sqrt{z}F(z),
\end{equation}
transforms the last equation to
\begin{equation}
	\frac{d^{2}{M}(z)}{{dz^{2}}}+\left[  -\frac{1}{4}-\frac{i\lambda}{z}%
	+\frac{\frac{1}{4}-\gamma_{l}^{2}}{z^{2}}\right]  M(z)=0, \label{32}%
\end{equation}
where $i\lambda=\frac{1}{2}-i\eta$ and $\eta=\frac{EV_{0}}{K}$.\newline This
equation admits two regular linearly independent solutions in terms of the
Whittaker functions \cite{22}
\begin{equation}
	u_{1}{(z)}={C_{3}}{M_{-i\lambda,{\gamma_{l}}}}(z)={C_{3}}{z^{({\gamma_{l}%
			}+\frac{1}{2})}}{e^{-z/2}}M({\gamma_{l}}+i\lambda+\frac{1}{2},2{\gamma_{l}%
	}+1;z), \label{33}%
\end{equation}%
\begin{equation}
	u_{2}{(z)}={C_{4}}{W_{-i\lambda,{\gamma_{l}}}}(z)={C_{4}}{z^{({\gamma_{l}%
			}+\frac{1}{2})}}{e^{-z/2}}U({\gamma_{l}}+i\lambda+\frac{1}{2},2{\gamma_{l}%
	}+1;z), \label{46}%
\end{equation}
where the first solution is bounded at $z=0$ while the second is bounded at
$\vert z \vert \rightarrow\infty$, $C_{3}$ and $C_{4}$ are the normalization constants.

\subsection{Creation of fermionic particles}
In order to obtain the expressions of the probability and the number density of created particles, we proceed as in the Klein-Gordon case using the formulas (\ref{f18}-\ref{224}) with ${k=i\lambda}$ and ${\mu={{{\gamma}_{l}=i}}}\tilde{\gamma_{l}}$. 

Then, the Bogoliubov coefficients $A$ and $B$ verify 

\begin{equation}
	\left\vert \frac{B}{A}\right\vert =\left\vert \frac{\Gamma(1-i(
		\tilde{\gamma_{l}}+\eta))}{\Gamma(1+i(\tilde{\gamma_{l}}-\eta))}
	e^{i\pi\left(  i\tilde{\gamma_{l}}-\frac{1}{2}\right)}\right\vert,
\end{equation}
and the probability of created fermionic particles is
\begin{equation}
	P=\left\vert \frac{B}{A}\right\vert ^{2}=\left\vert  \frac{\Gamma(1-i(
		\tilde{\gamma_{l}}+\eta))}{\Gamma(1+i(\tilde{\gamma_{l}}-\eta))}\right\vert
	^{2}e^{-2\pi\tilde{\gamma_{l}}},
\end{equation}
where%
\begin{equation}
	\tilde{\gamma_{l}}=\sqrt{V_{0}^{2}-\left[  {j+\left\vert m\right\vert \left(
			{\frac{1}{\alpha}-1}\right)  +\frac{1}{2}}\right]  ^{2}}>0, \label{cond2}%
\end{equation}
is the condition for pair production of spin-1/2 particle which depends on
${V_{0}}={qQ_{s}}$, the cosmic string parameter $\alpha$ and the quantum
numbers $j$ and $m$.

Using the fermionic condition $\left\vert A\right\vert ^{2}+\left\vert
B\right\vert ^{2}=1$ and the following property of the gamma function
\begin{equation}
	{\left\vert {\Gamma(1+ix)}\right\vert ^{2}}=\frac{\pi x}{\sinh(\pi x)}, 
\end{equation}
the expression of the probability is
\begin{equation}
	P=\frac{(\tilde{\gamma_{l}}+\eta)}{(\tilde{\gamma_{l}}-\eta)}\frac
	{\sinh\left[  \pi(\tilde{\gamma_{l}}-\eta)\right]  }{\sinh\left[  \pi
		(\tilde{\gamma_{l}}+\eta)\right]  }e^{-2\pi\tilde{\gamma_{l}}}. \label{633}%
\end{equation}
Then, the number density of the created fermionic particles is calculated as
\begin{equation}
	\bar{n}=\frac{1}{P^{-1}+1}=\frac{(\tilde{\gamma_{l}}+\eta)\sinh\left[
		\pi(\tilde{\gamma_{l}}-\eta)\right]  }{(\tilde{\gamma_{l}}-\eta)\sinh\left[
		\pi(\tilde{\gamma_{l}}+\eta)\right]  e^{2\pi\tilde{\gamma_{l}}}+(\tilde
		{\gamma_{l}}+\eta)\sinh\left[  \pi(\tilde{\gamma_{l}}-\eta)\right]  }.
	\label{D1}%
\end{equation}
At a large frequencies $(E\gg M)$, the probability (\ref{633})
reduces to
\begin{equation}
	P=\frac{(\tilde{\gamma_{l}}+V_{0})}{(\tilde{\gamma_{l}}-V_{0})}\frac
	{\sinh\left[  \pi(\tilde{\gamma_{l}}-V_{0})\right]  }{\sinh\left[  \pi
		(\tilde{\gamma_{l}}+V_{0})\right]  }e^{-2\pi\tilde{\gamma_{l}}}. \label{63}%
\end{equation}
and the number of created particle (\ref{D1}) can be written as
\begin{equation}
	\bar{n}=\frac{(\tilde{\gamma_{l}}+V_{0})\sinh\left[  \pi(\tilde{\gamma_{l}%
		}-V_{0})\right]  }{(\tilde{\gamma_{l}}-V_{0})\sinh\left[  \pi(\tilde
		{\gamma_{l}}+V_{0})\right]  e^{2\pi\tilde{\gamma_{l}}}+(\tilde{\gamma_{l}%
		}+V_{0})\sinh\left[  \pi(\tilde{\gamma_{l}}-V_{0})\right]  }. \label{D2}%
\end{equation}
which is not a thermal distribution.

For the case when  $\left\vert\tilde{\gamma_{l}}\pm V_{0}\right\vert\gg1$, the number
density takes the form of Fermi-Dirac distribution
\begin{equation}
	\bar{n}=\frac{1}{e^{2\pi(\tilde{\gamma}_{l}-V_{0})}+1}. \label{FD}%
\end{equation}
In the absence of the cosmic string $(\alpha=1)$, or for $m=0,$ the
probability (\ref{63}) and the number density (\ref{D2})  reduce to
\begin{equation}
	P(\alpha=1)=\frac{(\tilde{\gamma}_{1}+V_{0})}{(\tilde{\gamma}_{1}-V_{0})}%
	\frac{\sinh\left[  \pi(\tilde{\gamma}_{1}-V_{0})\right]  }{\sinh\left[
		\pi(\tilde{\gamma}_{1}+V_{0})\right]  }e^{-2\pi\tilde{\gamma}_{1}},
\end{equation}
\begin{equation}
	\bar{n}(\alpha=1)=\frac{(\tilde{\gamma}_{1}+V_{0})\sinh\left[  \pi
		(\tilde{\gamma}_{1}-V_{0})\right]  }{(\tilde{\gamma}_{1}-V_{0})\sinh\left[
		\pi(\tilde{\gamma}_{1}+V_{0})\right]  e^{2\pi\tilde{\gamma}_{1}}%
		+(\tilde{\gamma}_{1}+V_{0})\sinh\left[  \pi(\tilde{\gamma}_{1}-V_{0})\right]
	}, \label{D22}%
\end{equation}
where $\tilde{\gamma}_{1}\equiv\tilde{\gamma}_{l}(\alpha=1)=\sqrt{V_{0}%
	^{2}-{{\left(  {j+\frac{1}{2}}\right)  }^{2}}}$.

\subsection{Discussion of results}

$\bullet$ First, let us discuss the consequences of the condition
$(\ref{cond2})$ for pair production of spin-1/2 fermion. It can be written as
\begin{equation}
	\left\vert qQ_{s}\right\vert > \left[  j+\left\vert m\right\vert \left(  \frac{1}%
	{\alpha}-1\right)  +\frac{1}{2}\right]  . \label{cond22}%
\end{equation}
In the absence of the cosmic string ($\alpha=1)$, or for $m=0$, the above
condition is reduced to
\begin{equation}
	\left\vert qQ_{s}\right\vert > \left(  j+\frac{1}{2}\right)  =\left(  l+1\right)  .
\end{equation}
For an electron in a hydrogen-like atom, $Q_{s}=Ze$ and $q=-e$, the 
condition (\ref{cond22}) can be expressed for the nucleus charge $Z$ as
\begin{equation}
	Z>Z_{cr}\equiv\frac{1}{e^{2}}\left[j+\left\vert m\right\vert (\frac{1}{\alpha
	}-1)+\frac{1}{2}\right]  .
\end{equation}
For the particular values of the quantum numbers $(l=0,m=0,s=1/2,j=1/2)$, the
pair production of fermionic particles is possible for $Z>Z_{cr}=\frac
{1}{e^{2}}=137$ (in units $\hbar=c=1$, $e^{2}\simeq1/137$), which is a well-known result in the absence of the cosmic string \cite{popov1}.

In table \ref{table2}, for the first values of the quantum numbers
$(l,m,s=1/2,j=l+1/2),$ we give the critical values $Z_{cr}$ as function
of the string parameter $\alpha$ and an estimation of the effect of the GUT
cosmic string on the values and the rate of $Z_{cr}^{GUT}$.

\begin{table}[h]
\begin{center}
	\tabcolsep=18pt
	\renewcommand\arraystretch{1.2}
		\begin{minipage}{\textwidth}
			\caption{$Z_{cr}$ in terms of $\alpha$, values and the rate of $Z_{cr}^{GUT}$ for fermion particle} \label{table2}
	\begin{tabular}[c]{|c|c|c|c|c|c|}\hline
		$l$ & $m$ & $j=l+\frac{1}{2}$ & $Z_{cr}$ &
		$Z_{cr}^{GUT}$ & $\frac{Z_{cr}^{GUT}-Z_{cr}(\alpha=1)}{Z_{cr}(\alpha=1)} $\\\hline
		$0$ & $0$ & $\frac{1}{2}$ & $137$ &  & \\\hline
		$1$ & $0$ & $\frac{3}{2}$ & $137\times2=274$ &  & \\\cline{2-6}
		& $\pm1$ & $\frac{3}{2}$ & $137\times\left(  \frac{1}{\alpha}+1\right)  $ &
		$274,000137$ & $5,000005\times10^{-5}\%$\\\hline
		& $0$ & $\frac{5}{2}$ & $137\times3=411$ &  & \\\cline{2-6}%
		$2$ & $\pm1$ & $\frac{5}{2}$ & $137\times\left(  \frac{1}{\alpha}+2\right)  $
		& $411,000137$ & $3,333333\times10^{-5}\%$\\\cline{2-6}
		& $\pm2$ & $\frac{5}{2}$ & $137\times\left(  \frac{2}{\alpha}+1\right)  $ &
		$411,000274$ & $6,666666\times10^{-5}\%$\\\hline
	\end{tabular}
\end{minipage}
\end{center}
\end{table}
\vspace*{1mm}
We note that for the sub-states with $m=0$, the critical value $Z_{cr}$
increases with $l$, depends on the spin value $(s=1/2)$ and is independent of
the cosmic string parameter $\alpha$ (Note that these sub-states are
equivalent to the case $\alpha=1$). For the sub-states for which $m\neq0$,
$Z_{cr}$ increases with $l$, depends linearly on the spin value $(s=1/2)$ and
inversely on the cosmic string parameter $\alpha$. For the GUT
cosmic string, the obtained critical values $Z_{cr}^{GUT}$ do not exist for the moment, the rate of $Z_{cr}^{GUT}$ is about $10^{-5}\%$ as compared with the case $\alpha=1$
(without cosmic string) and the production of spin-1/2 particles is possible if the Coulomb potential nucleus charge $Z \geq275$. \\

\begin{figure}[h]
	\begin{center}%
		\begin{tabular}
			[c]{cc}%
			\includegraphics[width=0.5\textwidth]{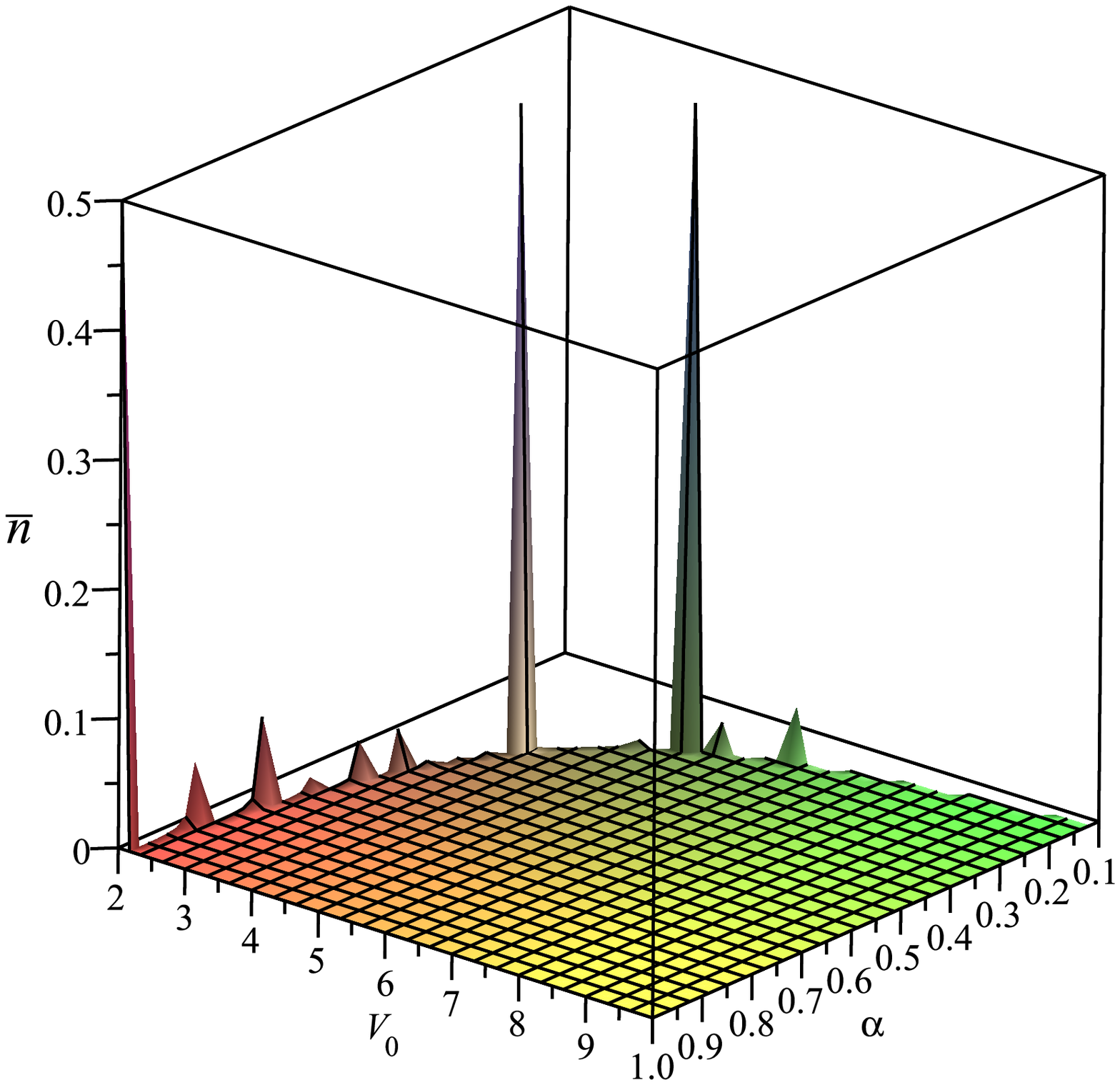} &
			\includegraphics[width=0.4\textwidth]{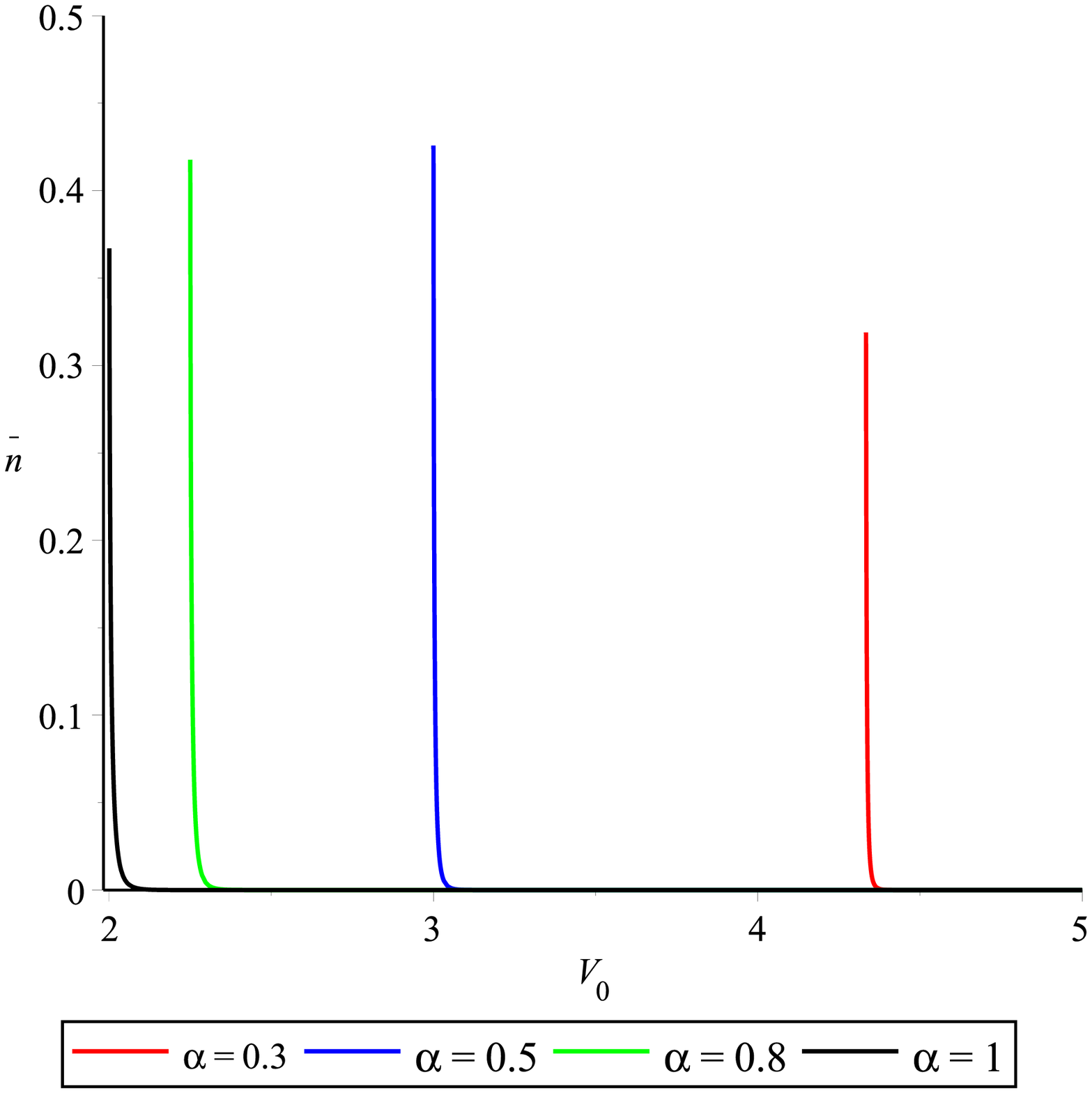}\\
			(a) & (b)
		\end{tabular}
	\end{center}
	\begin{center}
		\caption{Plot (a) is the 3D number density (\ref{D2}) in terms of $V_{0}\in\lbrack0,+10]$ and $\alpha\in\rbrack0,1]$. Plot (b) is the 2D number density (\ref{D2}) in terms of $V_{0}\in\lbrack0,5]$ for the different values of $\alpha=0.3,0.5,0.8$ and $1$ (without cosmic string) \label{fig2}.}
	\end{center}
\end{figure}

$\bullet$ Secondly, in Figure \ref{fig2} the number density curves have been plotted
for fixed values of the quantum numbers $(l=1,m=1,s=1/2,j=3/2)$ and where the condition $(\ref{cond2})$ for pair production of 
spin-1/2 particles is reduced to $V_{0} > (1+\frac{1}{\alpha})$ which should be always fullfilled. Plot (a) display, in three (3D) dimensions, the number density $(\ref{D2})$ in terms of $V_{0}=qQ_{s}$ and the cosmic string parameter $\alpha$. Plot (b) display, in two (2D) dimensions, the number density $(\ref{D2})$ in terms of $V_{0}=qQ_{s}$ for different values of the cosmic string parameter $\alpha$.

\begin{figure}[h]
	\begin{center}
		\includegraphics[width=0.5\textwidth]{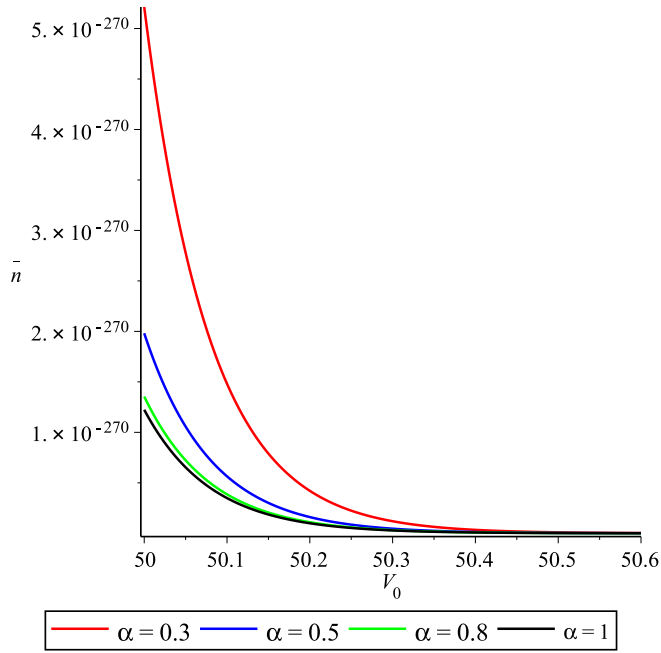}
		\caption{Number density (\ref{D2}) in terms of $V_{0}\in\lbrack50.0,50.6]$ for different values of $\alpha=0.3, 0.5, 0.8$ and $1$ (without cosmic string) \label{fig3}.}
	\end{center}
\end{figure}

In order to study the shape of the curves for large values of $V_{0}$, the number density (\ref{D2}) has been plotted in figure \ref{fig3} in the
short interval of $V_{0}\in\lbrack50.0,50.6]$ for different values of $\alpha$. From which we note that the number density $\bar{n}$ decreases when the cosmic string parameter $\alpha$ increases (i.e. the linear mass density $\mu$ of the string decreases) where its smallest values correspond to the case $\alpha = 1$ (in the absence of the cosmic string). Therefore, we deduce that the presence of the cosmic string improves the number density of fermion particles compared to the case without the cosmic string ($\alpha=1$).

\section{Conclusion}

In this paper, we have investigated the influence of a cosmic string on the pair production rate induced by the Coulomb potential of an external electric field. The solutions of the radial Klein-Gordon and Dirac equations are given in terms of Whittaker functions,
the probability and the number density of created particles have been calculated.

For the spin-$0$ boson case, the creation of particles holds for 
$\left\vert qQ_{s}\right\vert > {{\left[l+\left\vert m\right\vert \left({\frac{1}{\alpha}-1}\right)  {+\frac{1}{2}}\right]}}$, the number density decreases when the cosmic string parameter $\alpha$ increases in the interval $]0,1]$ and its smallest values are reached for $\alpha=1$. Thus, the presence of the cosmic string improves the number density of created spinless particles compared to the case without the cosmic string ($\alpha=1$). For the particular values of the quantum numbers $(l=0,m=0)$, the pair production of scalar particles is possible if the Coulomb potential nucleus charge $Z>\frac{137}{2}$.

For the spin-$1/2$ fermion case, the creation of particles holds for 
$ \left\vert qQ_{s} \right\vert > \left[{j+\left\vert m\right\vert \left({\frac{1}{\alpha}-1}\right)+\frac{1}{2}}\right]$, the number density decreases with the increasing of the cosmic string parameter $\alpha$ in the interval $]0,1]$ and its smallest values are reached for $\alpha=1$. Thus, the presence of the cosmic string improves also the number density of created spin-1/2 particles compared to the case without the cosmic string ($\alpha=1$). For the particular values of the quantum numbers $(l=0,m=0,s=1/2,j=1/2)$, the pair production of fermionic particles is possible if the Coulomb potential nucleus charge $Z>137$.

As a result for the GUT cosmic string, the pair production for spinless particles is possible if the Coulomb potential nucleus charge $Z\geq206$, and for spin-1/2 particles if $Z\geq275$.

In limiting case of the Minkowski spacetime ($\alpha=1$), as expected we
retrieve the expressions of the number density of the scalar and fermionic
particles created by the Coulomb potential in the absence of the cosmic
string. Finally, we note that the results for the sub-states for which the magnetic quantum number
$m=0$ are independent of the cosmic string parameter $\alpha$,  i.e. these sub-states are insensitive to the presence of the cosmic string.

\end{document}